\documentclass[aps,prb,twocolumn,floats]{revtex4}
\usepackage{graphics}
\usepackage{amssymb}
\usepackage{amsmath}
\usepackage{bm}
\usepackage{color}

\begin{document}

\title{Piecewise-terminated spherical topological insulator as a virtual breadboard for Majorana circuitry}
\author{Adam C. Durst$^1$ and Sriram Ganeshan$^2$}
\affiliation{$^1$Department of Physics and Astronomy, Hofstra University, Hempstead, NY 11549-1510, USA}
\affiliation{$^2$Department of Physics, The City College of New York, New York, NY 10031, USA}

\date{August 6, 2018}

\begin{abstract}
We consider the surface states of a spherical topological insulator piecewise-terminated by superconductivity or ferromagnetism over various regions of the spherical surface.  Such terminations gap the surface states by breaking U(1) particle-number symmetry or time-reversal symmetry, respectively.  Interfaces and trijunctions between differently terminated surface regions can host propagating and bound Majorana modes, and the finite size of the spherical system makes it easily amenable to numerical analysis via exact diagonalization of the Bogoliubov-de Gennes Hamiltonian within a truncated Hilbert space.  Creative termination patterning therefore allows one to prototype a variety of Majorana circuits, calculating energy spectra and plotting eigenfunctions over the spherical surface.  We develop the computational framework for this approach, establishing a virtual breadboard for Majorana circuitry, and apply it to circuits of interest, including the Majorana analog of a Mach-Zehnder interferometer.
\end{abstract}

\maketitle

\section{Introduction}
\label{sec:intro}

The study of symmetry protected topological phenomena~\cite{moore2010birth, hasan2010colloquium, qi2011topological, bernevig2013topological} has experienced tremendous growth over the past decade with the theoretical prediction~\cite{fu2007topological, moore2007topological, fu2007topologicalprb, roy2009topological} of three-dimensional topological insulators (TIs) followed by their experimental realization~\cite{zhang2009topological, chen2009experimental, xia2009observation, hsieh2009observation}. The interplay of topological protection and symmetries has enlarged the library of topological phenomena and may become the foundation of next-generation quantum technologies~\cite{kitaev2003fault, nayak2008non, hasan2010colloquium, qi2011topological, bernevig2013topological}. One of the hallmarks of a three-dimensional topological insulator is the presence of gapless surface states that are protected by time reversal and charge conservation symmetry~\cite{xia2009observation, hsieh2009observation}. These surface states can be gapped by breaking the time reversal symmetry, either by application of magnetic field or by coating with a magnetic insulator, leading to a novel quantum Hall effect at the surface. Apart from breaking time-reversal symmetry, one can also gap the surface states by breaking the charge conservation through proximity to a superconductor. The existence of multiple gapping mechanisms (magnetic and superconducting) allows us to create different types of domain walls that are interfaces of two independent gapping terms. These domain wall interfaces can host propagating or bound-state neutral Majorana excitations that can be leveraged to create topologically protected Majorana circuits~\cite{fu2008superconducting, fu2009probing}. Some examples of these Majorana circuits include analogs of Mach-Zehnder and Fabry-Perot interferometers~\cite{fu2009probing, akhmerov2009electrically}. Such circuits act as experimental probes capable of detecting Majorana excitations as well as assisting in controlling and manipulating them for technological applications.

Since these circuits of Majorana excitations live exclusively on the TI surface, it is important to study TI surface states in the presence of realistic material effects such as disorder, proximity effects, and interactions.  However, numerical analysis of such effects has proved challenging due to the fermion-doubling theorem~\cite{nielsen1981no} prohibiting a lattice description.  This is because the surface states can only exist on the boundary of some higher-dimensional topological phase.  Recent works~\cite{imura2012spherical, neupert2015interacting, greiter2018landau} have attempted to overcome this limitation by developing a continuum theory of a single Dirac cone on the surface of a topological insulator of spherical geometry.

In this paper, we build upon such efforts by implementing Majorana circuits on the surface of a spherical topological insulator. Chiral Majorana channels are realized at the interface of surface regions with different gapping terms.  Such interfaces act as wires for the propagation of Majorana modes.  In order to include regions gapped via proximity-induced superconductivity, we have extended the approach of Neupert {\it et al.\/} \cite{neupert2015interacting} to a formalism based on the Bogoliubov-de Gennes (BdG) equation.  The spherical manifold allows us to study probe circuits such as a Mach-Zehnder interferometer, and also enables us to enclose an additional bound-state Majorana acting as a $\pi$-flux within the area of the interferometer. There are several technical advantages to studying these circuits numerically in the spherical geometry. The finite radius of the sphere results in a well-regulated discrete energy spectrum with well-defined eigenstates that can be labeled by good quantum numbers coming from the spherical symmetry. The time-reversal symmetry of the spherical TI surface can be preserved by the introduction of a fictitious magnetic monopole~\cite{imura2012spherical} at the center of the sphere that has opposite sign for electrons versus holes. Imura {\it et al.\/}~\cite{imura2012spherical} showed that the large radius limit reproduces flat manifold physics, which enables us to apply our results to realistic experimental setups.

This paper is organized as follows. In Sec.~\ref{sec:formulation}, we develop the formulation for our numerical calculation.  Starting with the Bogoliubov-de Gennes equation appropriate to the spherical TI surface, we establish basis states, define a framework for inputting surface termination patterns, compute matrix elements, calculate energy spectra and eigenstates, and derive expressions for quasiparticle density and quasiparticle current density.  In Sec.~\ref{sec:results}, we demonstrate the utility of our numerics by implementing a series of termination patterns and studying the resulting Majorana circuits, from propagating equatorial modes, to polar bound states, to a Mach-Zehnder interferometer.  In Sec.~\ref{sec:conclusions}, we summarize our results and outline future directions.

\section{Formulation}
\label{sec:formulation}
Consider a spherical topological insulator (TI) terminated by either superconductivity or ferromagnetism over different regions of the spherical surface.  We seek solutions to the Bogoliubov-de Gennes (BdG) equation for the resulting surface, characterized by a piecewise-uniform proximity-induced superconducting order parameter $\Delta$ and out-of-surface magnetization $M$.  The $\Delta=M=0$ case can be solved exactly~\cite{neupert2015interacting}, with solutions labeled by quantum numbers $n$, $m$, $\lambda$, and $\gamma$.  Such solutions provide a basis for our Hilbert space, which can be truncated~\cite{neupert2015interacting} by restricting to states with $n \leq n_{\rm max}$.  For a given piecewise-uniform termination pattern (a map of regions with nonzero $\Delta$ or $M$ over the surface of the sphere) our approach is to evaluate matrix elements of the BdG Hamiltonian within our basis and diagonalize to obtain energy spectra and eigenstates.  Quasiparticle density and quasiparticle current density for each eigenstate can then be evaluated and plotted over the surface of the sphere.  Such solutions are exact within the restricted Hilbert space, with larger $n_{\rm max}$ corresponding to higher angular resolution.  In this work, we have limited ourselves to termination patterns with regions bounded by segments of the equator and/or lines of longitude (meridians). Doing so allows for matrix elements to be computed very efficiently, with minimal quadrature, and in turn, makes it computationally feasible to include basis states up to $n_{\rm max}=10$ (a 528-dimensional Hilbert space).

\subsection{Dirac BdG Hamiltonian}
\label{ssec:DBdG}
The Bogoliubov-de Gennes (BdG) equation \cite{degennes1966superconductivity} has the following general form \cite{beenakker2006specular, leijnse2012introduction}
\begin{equation}
H_{\rm BdG} \Psi = E \Psi
\label{eq:BdGequation}
\end{equation}
\begin{equation}
H_{\rm BdG} = \left[ \begin{array}{cc} H - \mu & \Delta \\ \Delta^\dagger & \mu - T H T^{-1} \end{array} \right]
\label{eq:HBdGdef}
\end{equation}
where $H$ is the single-particle Hamiltonian, $\mu$ is the chemical potential, $\Delta$ is the proximity-induced superconducting order parameter, and $T$ is the time-reversal operator.  Note that $\Psi$ is a 4-component Nambu spinor and $H_{\rm BdG}$ is a $4 \times 4$ matrix since $H$, $\Delta$, and $T$ are all, in general, $2 \times 2$ matrices acting on spin space.  In what follows, we make use of two sets of Pauli matrices, $\sigma_i$ acting on spin space and $\tau_i$ acting on particle-hole space.  The coordinate system for our spins is defined locally over the spherical surface, with $\sigma_1$, $\sigma_2$, and $\sigma_3$ referring to the $\hat{\bm{\phi}}$ (east), $-\hat{\bm{\theta}}$ (north), and $\hat{\bf r}$ (out-of-surface) directions respectively.  With these definitions, $T = -i \sigma_2 K$ (where $K$ denotes complex conjugation) and
\begin{equation}
H = H_0 + {\bf M} \cdot {\bf \sigma}
\label{eq:Hsingleparticle}
\end{equation}
where ${\bf M}$ is the proximity-induced magnetization vector (in units of energy).  Here $H_0$ is the Hamiltonian of the bare TI surface, a two-dimensional massless Dirac Hamiltonian in spherical geometry, which was shown by Imura {\it et al.\/} \cite{imura2012spherical} to take the form
\begin{equation}
H_0 = \frac{v}{R} \left( \sigma_1 \Lambda_\theta + \sigma_2 \Lambda_\phi \right)
\label{eq:sphericalDirac}
\end{equation}
\begin{equation}
{\bf \Lambda} = -i \left[ \hat{\bf \phi} \frac{\partial}{\partial\theta}
- \hat{\bf \theta} \frac{1}{\sin\theta} \left( \frac{\partial}{\partial\phi} - \frac{i}{2} \cos\theta \sigma_3 \right) \right]
\label{eq:monopole}
\end{equation}
where $v$ is the slope of the Dirac cone and $R$ is the radius of the spherical surface.
Plugging Eq.~(\ref{eq:Hsingleparticle}) into Eq.~(\ref{eq:HBdGdef}) yields
\begin{equation}
H_{\rm BdG} = \left[ \begin{array}{cc} H_0 + {\bf M} \cdot {\bf \sigma} - \mu & \Delta \\
\Delta^\dagger & \mu - H_0 + {\bf M} \cdot {\bf \sigma} \end{array} \right]
\label{eq:HBdGgeneral}
\end{equation}
because
\begin{eqnarray}
T H_0 T^{-1} &=& H_0 \nonumber \\
T {\bf M} \cdot {\bf \sigma} T^{-1} &=& -{\bf M} \cdot {\bf \sigma}
\label{eq:TR}
\end{eqnarray}
as long as ${\bf M}$ is a real function of angles $\theta$ and $\phi$.  In this work, we consider only $s$-wave superconductivity (scalar $\Delta$) and out-of-surface magnetization (${\bf M} = M \hat{\bf r}$), so
\begin{equation}
H_{\rm BdG} = \left[ \begin{array}{cc} H_0 + M\sigma_3 - \mu & \Delta \\
\Delta^* & \mu - H_0 + M\sigma_3 \end{array} \right],
\label{eq:HBdG}
\end{equation}
but the formalism and numerics can be easily adapted to other situations.

\subsection{Basis states}
\label{ssec:basis}
The single-particle Hamiltonian, $H_0$, of the unterminated spherical TI surface yields exact solutions indexed by quantum numbers $n$, $m$, and $\lambda$, where $n$ is a nonnegative integer, $m=-s,-s+1,...,s$ (where $s=n+\frac{1}{2}$), and $\lambda=\pm 1$.  Neupert {\it et al.\/} \cite{neupert2015interacting}, making use of formalism developed in Refs.~\onlinecite{greiter2011landau} and \onlinecite{haldane1983fractional}, showed that
\begin{equation}
H_0 \psi_{nm}^\lambda = \epsilon_{nm}^\lambda \psi_{nm}^\lambda
\label{eq:H0equation}
\end{equation}
where the $2(n+1)$-fold degenerate eigenvalues are
\begin{equation}
\epsilon_{nm}^\lambda = \lambda \left( n + 1 \right) \frac{v}{R}
\label{eq:H0eigenvalue}
\end{equation}
and the spinor eigenstates take the form
\begin{equation}
\psi_{nm}^\lambda (\theta,\phi)
= \left( \begin{array}{c}
\phi_{nm}^\uparrow \\
\lambda \phi_{nm}^\downarrow \end{array} \right)
\label{eq:Psinml}
\end{equation}
with
\begin{equation}
\phi_{nm}^\uparrow = (L^-)^{s-m} \bar{v}^n u^{n+1}
\;\;\;\;\;\;\;\;
\phi_{nm}^\downarrow = -\frac{S^-}{n+1} \phi_{nm}^\uparrow
\label{eq:phiupdown}
\end{equation}
where the $L^-$ and $S^-$ operators
\begin{equation}
L^- \equiv v \partial_u - \bar{u} \partial_{\bar{v}}
\;\;\;\;\;\;\;\;
S^- \equiv \bar{v} \partial_u - \bar{u} \partial_v
\label{eq:LminusSminus}
\end{equation}
are defined in terms of the spinor coordinates
\begin{equation}
u \equiv \cos(\theta/2) e^{i\phi/2}
\;\;\;\;\;\;\;\;
v \equiv \sin(\theta/2) e^{-i\phi/2} .
\label{eq:uvdef}
\end{equation}
We can therefore write down solutions to the Bogoliubov-de Gennes equation for the unterminated case where $\Delta = M = 0$.  In this case,
\begin{equation}
H_{\rm BdG}^0 = \left[ \begin{array}{cc} H_0 - \mu & 0 \\
0 & \mu - H_0 \end{array} \right]
\label{eq:HBdG0}
\end{equation}
and
\begin{equation}
H_{\rm BdG}^0 \Psi_{0\,nm}^{\lambda\gamma} = E_{0\,nm}^{\lambda\gamma} \Psi_{0\,nm}^{\lambda\gamma}
\label{eq:HBdG0equation}
\end{equation}
where we have introduced a fourth quantum number, $\gamma = \pm 1$.  The bare (unterminated) energy spectrum takes the form
\begin{equation}
E_{0\,nm}^{\lambda\gamma} = \gamma \left( \epsilon_{nm}^\lambda - \mu \right)
\label{eq:E0}
\end{equation}
with bare (unterminated) eigenstates
\begin{equation}
\Psi_{0\,nm}^{\lambda\gamma} = \left\{ \begin{array}{r}
\left[ \begin{array}{c} \psi_{nm}^\lambda \\ 0 \end{array} \right] \vspace{0.2cm} \;\;\;\; {\rm for} \;\;\;\; \gamma = +1\\
\left[ \begin{array}{c} 0 \\ T \psi_{nm}^\lambda \end{array} \right] \;\;\;\; {\rm for} \;\;\;\; \gamma = -1
\end{array} \right.
\label{eq:Psi0}
\end{equation}
where the zeros denote two-component spinor zeros.  Such eigenstates are orthogonal and we normalize over the unit sphere.

These bare eigenstates provide the basis that we will use to study the full (piecewise terminated by $\Delta$ or $M$) BdG Hamiltonian.  In principle, since $n$ is unbounded from above, there are an infinite number of them.  However, as discussed in Refs.~\onlinecite{neupert2015interacting} and \onlinecite{durst2016disorder}, the bulk band gap provides a natural cutoff, $\Lambda$, for the single-particle energy spectrum.  Thus, following Refs.~\onlinecite{neupert2015interacting} and \onlinecite{durst2016disorder}, we truncate our Hilbert space at $n = n_{\rm max} \equiv  \lfloor \frac{R\Lambda}{v} - 1 \rfloor$, where the brackets denote the greatest integer or floor function.  Doing so limits the number of single-particle states to $N = 2(n_{\rm max} + 1)(n_{\rm max} + 2)$, such that there are $2N$ states in the BdG problem.  Since a principal aim of our work is to use the spherical model as a computational aid to study the flat TI surface, we are primarily interested in the large $R$, and therefore large $n_{\rm max}$, limit.  From this point of view, Hilbert space truncation at finite $n_{\rm max}$ is a computational approximation that serves to limit the angular resolution of our results.  Trading computation time (which increases quickly with increasing $n_{\rm max}$) versus angular resolution, we work with $n_{\rm max}=10$, which yields 528 BdG states.

Inspection of Eqs.~(\ref{eq:Psinml}) through (\ref{eq:uvdef}) reveals that each of the four components of each of our basis states, $\Psi_{0\,nm}^{\lambda\gamma}$, can be expressed as a polynomial of the form
\begin{equation}
\sum_{k=1}^{N_{\rm SCE}} A_k \left( \sin \frac{\theta}{2} \right)^{p_k} \left( \cos \frac{\theta}{2} \right)^{q_k} \left( e^{i\phi/2} \right)^{r_k}
\label{eq:SCEpoly}
\end{equation}
where the $A_k$ are complex numbers and the $p_k$, $q_k$, and $r_k$ are integers.  Such polynomials were dubbed SCE (sin-cos-exp) polynomials in Ref.~\onlinecite{durst2016disorder}.  Thus, in our numerics, it is convenient to store each basis state component as a $4 \times N_{\rm SCE}$ matrix of $A$-$p$-$q$-$r$ parameters, where $N_{\rm SCE}$ is the number of terms in that SCE polynomial.  Since the set of SCE polynomials is closed under addition, multiplication, and complex conjugation, such data structures are easily manipulated and (as shown in the next section) efficiently integrated to compute matrix elements.

\subsection{Matrix elements}
\label{ssec:matrixelements}
With basis states in hand, we proceed to evaluate the matrix elements of the full BdG Hamiltonian, $H_{\rm BdG}$, in the basis of the eigenstates $\Psi_{0\,nm}^{\lambda\gamma}$ of the bare (unterminated) BdG Hamiltonian, $H_{\rm BdG}^0$.  If $i$ and $j$ index the $2N$ basis states, these matrix elements take the form
\begin{eqnarray}
H_{ij} &\equiv& \left\langle \Psi_{0\,n_i m_i}^{\lambda_i \gamma_i} \left| H_{\rm BdG} \right| \Psi_{0\,n_j m_j}^{\lambda_j \gamma_j} \right\rangle \nonumber \\
&=& E_{0\,n_i m_i}^{\lambda_i \gamma_i} \delta_{ij} + \Delta H_{ij}
\label{matrixelements}
\end{eqnarray}
where
\begin{equation}
\Delta H_{ij} = \left\langle \Psi_{0\,n_i m_i}^{\lambda_i \gamma_i} \left| \Delta H_{\rm BdG} \right| \Psi_{0\,n_j m_j}^{\lambda_j \gamma_j} \right\rangle
\label{eq:DeltaHij}
\end{equation}
and
\begin{equation}
\Delta H_{\rm BdG} = \left[ \begin{array}{cc} M \sigma_3 & \Delta \\
\Delta^* & M \sigma_3 \end{array} \right] .
\label{eq:DeltaHBdG}
\end{equation}
If $M$ and $\Delta$ are general functions of $\theta$ and $\phi$, the 2D integral over the unit sphere in Eq.~(\ref{eq:DeltaHij}) must be computed numerically.  But if $M$ and $\Delta$ are piecewise-uniform, then for each region of uniform $M$ and $\Delta$, the integrand is itself an SCE polynomial (since sums and products of SCE polynomials are SCE polynomials), and such integrals take the form
\begin{equation}
\Delta H_{ij}^{\rm region} = \sum_k A_k \iint \!\!d\theta d\phi \left( \sin \frac{\theta}{2} \right)^{p_k} \!\!\left( \cos \frac{\theta}{2} \right)^{q_k} \!\!\left( e^{i\phi/2} \right)^{r_k}
\label{eq:IntegralSCE}
\end{equation}
where the $A$-$p$-$q$-$r$ constants are those appropriate to the entire integrand, including the $\sin\theta$ measure.  If we specify that all boundaries between regions must be segments of lines of longitude or latitude, then Eq.~(\ref{eq:IntegralSCE}) becomes
\begin{equation}
\Delta H_{ij}^{\rm region} = \sum_k A_k \,I_\theta(p_k,q_k,\theta_1,\theta_2) \,I_\phi(r_k,\phi_1,\phi_2)
\label{eq:IntegralSCESpecialRegions}
\end{equation}
where
\begin{equation}
I_\theta(p_k,q_k,\theta_1,\theta_2) = \int_{\theta_1}^{\theta_2} \left( \sin \frac{\theta}{2} \right)^{p_k} \!\!\left( \cos \frac{\theta}{2} \right)^{q_k} d\theta
\label{eq:Itheta}
\end{equation}
\begin{eqnarray}
I_\phi(r_k,\phi_1,\phi_2) &=& \int_{\phi_1}^{\phi_2} e^{ir_k\phi/2} d\phi \\
&=& \left\{ \begin{array}{r}
\frac{2}{ik} \left( e^{ir_k\phi_2/2} - e^{ir_k\phi_1/2} \right) \vspace{0.2cm} \;\;\; {\rm for} \;\;\; r_k \neq 0\\
\phi_2 - \phi_1 \;\;\;\;\;\;\;\;\;\; \;\;\; {\rm for} \;\;\; r_k = 0
\end{array} \right. \nonumber
\label{eq:Iphi}
\end{eqnarray}
and we can take advantage of the fact that the $I_\phi$ integral is easily evaluated analytically.

In the present work, we consider only termination patterns with regions bounded by segments of meridians (lines of longitude) and the equator.  Thus, we need only evaluate $I_\theta$ from $0$ to $\pi/2$ and from $\pi/2$ to $\pi$.  It is straightforward to show that the latter integral is equal to the former with $p$ and $q$ indices interchanged.
\begin{equation}
I_S(p,q) \equiv I_\theta(p,q,\pi/2,\pi) = I_\theta(q,p,0,\pi/2) \equiv I_N(q,p)
\label{eq:northsouth}
\end{equation}
Thus, computation of matrix elements reduces to the numerical evaluation of $I_N(p,q)$ for a limited set of whole number indices $p$ and $q$.  In practice, we pre-compute these once for all required indices and save to a look-up table, eliminating the need to do any numerical integration on the fly, which vastly speeds up the computation of matrix elements and allows us to maximize the dimensionality of our Hilbert space.

\subsection{Exact diagonalization}
\label{ssec:diagonalization}
Matrix element computation yields the $2N \times 2N$ matrix, $H_{ij}$.  In the present work, with $n_{\rm max}=10$, this is a $528 \times 528$ matrix.  We diagonalize it numerically to obtain the $2N$ eigenvalues and eigenvectors.  The eigenvalues, $E_j$, are the energy levels of $H_{\rm BdG}$.  The $2N$ components of eigenvector ${\bf v}^j$ are the coefficients that define eigenstate $\Psi^j$ as a linear combination of the $2N$ bare eigenstates.
\begin{equation}
\Psi^j = \sum_i v_i^j \Psi_{0\,n_i m_i}^{\lambda_i \gamma_i}
\label{eq:EigenstateFromEigenvector}
\end{equation}
Each of the $2N$ eigenstates computed in this manner is a four-component Nambu spinor, where each component is a function of $\theta$ and $\phi$ that can be expressed as an SCE polynomial (see Eq.~(\ref{eq:SCEpoly})). Via Eq.~(\ref{eq:EigenstateFromEigenvector}), we compute the $A$-$p$-$q$-$r$ constants associated with each, and store all eigenstates in this convenient SCE format.

\subsection{Quasiparticle current density}
\label{ssec:current}
Since quasiparticle density is conserved, the quasiparticle current density functional, ${\bf j}[\Psi]$, is obtained from the quasiparticle density functional
\begin{equation}
\rho[\Psi] = \Psi^\dagger \Psi
\label{eq:density}
\end{equation}
via the continuity equation
\begin{equation}
\frac{\partial \rho}{\partial t} + {\bm \nabla} \cdot {\bf j} = 0 .
\label{eq:continuity}
\end{equation}
where $\Psi$ is the wave function, a four-component Nambu spinor.  Plugging in for $\rho$ and noting that $i\partial \Psi / \partial t = H_{\rm BdG}\Psi$, yields
\begin{equation}
i{\bm \nabla} \cdot {\bf j} = \left( H_{\rm BdG} \Psi \right)^\dagger \Psi - \Psi^\dagger \left( H_{\rm BdG} \Psi \right) .
\label{eq:divj1}
\end{equation}
which becomes
\begin{equation}
{\bm \nabla} \cdot {\bf j} = \frac{v}{R\sin\theta} \left[ \frac{\partial}{\partial\theta} \left( \sin\theta \Psi^\dagger \sigma_2 \tau_3 \Psi \right)
+ \frac{\partial}{\partial\phi} \left( -\Psi^\dagger \sigma_1 \tau_3 \Psi \right) \right]
\label{eq:divj2}
\end{equation}
once we have inserted $H_{\rm BdG}$ via Eqs.~(\ref{eq:sphericalDirac})-(\ref{eq:HBdGgeneral}).  Noting that the divergence of a surface vector in spherical coordinates \cite{jackson1975classical} has the form
\begin{equation}
{\bm \nabla} \cdot {\bf A} = \frac{1}{R\sin\theta} \left[ \frac{\partial}{\partial\theta} \left( \sin\theta A_\theta \right) + \frac{\partial A_\phi}{\partial\phi} \right] ,
\label{eq:divergence}
\end{equation}
this reduces to
\begin{equation}
{\bm \nabla} \cdot {\bf j} = {\bm \nabla} \cdot v \Psi^\dagger \left( \sigma_2 \hat{\bm \theta} - \sigma_1 \hat{\bm \phi} \right) \tau_3 \Psi
\label{eq:divj3}
\end{equation}
Thus, the quasiparticle current density functional is simply
\begin{equation}
{\bf j} = j_1 \hat{\bm \phi} + j_2 ( -\hat{\bm \theta} )
\label{eq:currentdensity}
\end{equation}
where the eastward component, $j_1$, and the northward component, $j_2$, take the form
\begin{equation}
j_1[\Psi] = -v \Psi^\dagger \sigma_1 \tau_3 \Psi \;\;\;\;\;\;\;\; j_2[\Psi] = -v \Psi^\dagger \sigma_2 \tau_3 \Psi .
\label{eq:j1j2}
\end{equation}
(Note that the minus signs are due to a sign convention, originally introduced in Ref.~\onlinecite{neupert2015interacting}, in the definition of the single-particle Dirac Hamiltonian.)  If the components of $\Psi$ are defined such that $\Psi^T = [\Psi_1, \Psi_2, \Psi_3, \Psi_4]$, then
\begin{equation}
j_1 = -2v\, {\rm Re} [\Psi_1^* \Psi_2 - \Psi_3^* \Psi_4]
\label{j1}
\end{equation}
\begin{equation}
j_2 = -2v\, {\rm Im} [\Psi_1^* \Psi_2 - \Psi_3^* \Psi_4] .
\label{j2}
\end{equation}
Once the $2N$ eigenstates, $\Psi^j$, have been computed as per Sec.~\ref{ssec:diagonalization}, it is straightforward to compute the quasiparticle density and quasiparticle current density for each, and then to plot these over the surface of the unit sphere.

\section{Results}
\label{sec:results}
We have developed a computational utility to apply the procedure described in Sec.~\ref{sec:formulation} to any piecewise-uniform $M$-$\Delta$ surface termination pattern with region boundaries that are either meridian segments or equator segments.  Here we explore the results of applying this utility to a number of simple termination patterns.  Throughout, we consider the case where the chemical potential is fixed at the Dirac point ($\mu = 0$), and we truncate our Hilbert space at $n_{\rm max} = 10$ (528 BdG states).

\subsection{Propagating Majorana modes}
\label{ssec:propagating}
\begin{figure}
\centerline{\resizebox{3.25in}{!}{\includegraphics{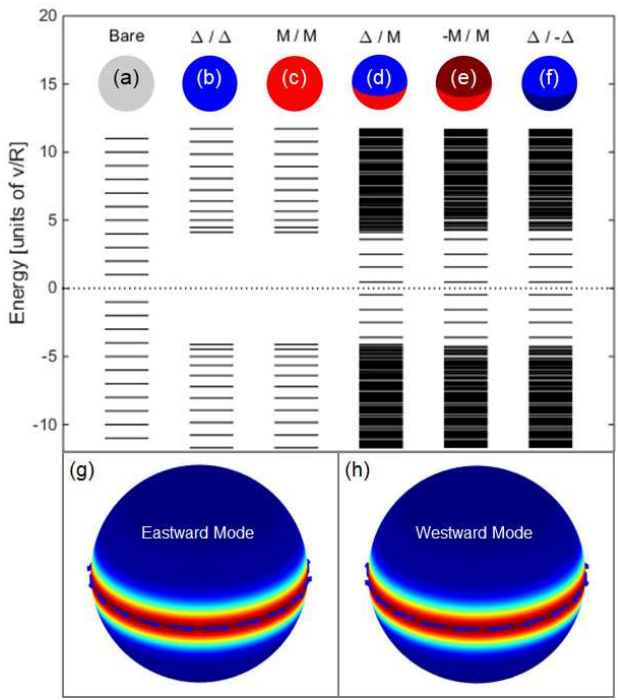}}}
\caption{Equatorial propagating Majorana modes.  (a)-(f) Energy spectra of the spherical TI surface terminated over the northern and southern hemispheres by various combinations of $s$-wave superconductor $\Delta$, out-of-surface magnetization $M$, and into-surface magnetization $-M$.  Plotted in (g) and (h) are the quasiparticle current density patterns corresponding to eastward and westward propagating equatorial Majorana modes (hue from blue to red denotes current magnitude, arrows denote current direction).  The lowest positive energy level of (d) yields a single eastward mode, of (e) yields two co-propagating eastward modes, and of (f) yields counter-propagating eastward and westward modes.  Results were computed via exact diagonalization up to $n_{\rm max} = 10$ (528 BdG states) for $|\Delta| = |M| = 4 v/R$ and $\mu = 0$.}
\label{fig:spectra}
\end{figure}

The simplest termination pattern, of course, is no termination at all.  In this case, the energy spectrum is given directly by Eqs.~(\ref{eq:E0}) and (\ref{eq:H0eigenvalue}).  For $\mu = 0$, the result is a ladder of energy levels of rung separation $v/R$, with the zero-energy rung absent and degeneracy increasing linearly away from zero energy (four times the rung number, counting away from zero in both the positive and negative directions).  Inputting no termination to our numerics yields this spectrum, trivially, as plotted in Fig.~\ref{fig:spectra}(a).

Terminating the entire sphere with an $s$-wave superconductor of order parameter $\Delta$ and constant phase breaks particle number conservation and opens up a gap from $-\Delta$ to $\Delta$.  Similarly, terminating the entire sphere with an out-of-surface magnetization $M$ breaks time reversal symmetry and opens up a gap from $-M$ to $M$.  Numerics-generated spectra for these two cases are plotted in Fig.~\ref{fig:spectra}(b) and (c), respectively.  In these cases and all that follow, we take $|\Delta| = |M| = 4 v/R$.

By applying different terminations to the northern and southern hemispheres, it is possible to introduce propagating interface states at the equator, with energies that fall within the gap.  (Several of these are discussed in figure 18.2 of Ref.~\onlinecite{bernevig2013topological}.  Here we demonstrate them with our numerics.)  A $\Delta$-terminated northern hemisphere with an $M$-terminated southern hemisphere (referred to in what follows as a $\Delta$/$M$ configuration) yields a chiral Majorana mode propagating eastward around the equator.  The resulting energy spectrum is shown in Fig.~\ref{fig:spectra}(d), where each within-gap energy level is nondegenerate.  The quasiparticle current density associated with the lowest positive energy state is plotted in Fig.~\ref{fig:spectra}(g).  Swapping $\Delta$ for $M$ or reversing the sign of the magnetization leaves the energy spectrum unchanged but changes the current direction from eastward to westward, as plotted in Fig.~\ref{fig:spectra}(h).  A $-M$/$M$ termination pattern yields two degenerate chiral Majorana modes propagating in the same direction (eastward) around the equator.  The energy spectrum is that of Fig.~\ref{fig:spectra}(e), where each within-gap energy level is two-fold degenerate.  Finally, a $\Delta$/$-\Delta$ termination pattern (an equatorial Josephson $\pi$-junction) yields two degenerate counterpropagating Majorana modes, one associated with an eastward equatorial current and the other associated with a westward equatorial current.  The energy spectrum computed for this configuration is that of Fig.~\ref{fig:spectra}(f), where each within-gap energy level is once again two-fold degenerate.

These simple two-region patterns provide helpful rules-of-thumb for understanding the propagating Majorana currents associated with more complex termination patterns.  The basic rules are as follows: (1) $\Delta$/$M$ and $\Delta$/$-M$ interfaces are single-lane one-way streets for Majorana current, oriented such that an $M$ region is to one's right, or a $-M$ region is to one's left, as one flows along with the current.  (2) $M$/$-M$ interfaces are two-lane one-way streets for Majorana current, hosting two co-propagating Majorana modes flowing such that the $M$ region is to one's right and the $-M$ region is to one's left as one flows along with the current.  Such interfaces can be thought of as $M$/$\Delta$/$-M$ double-interfaces, with an infinitely-narrow $\Delta$-median separating lanes.  (3) Josephson junction interfaces of the form $\Delta_1/\Delta_2$ only support propagating Majorana modes when their phase difference is $\pi$.  Such $\pi$-junction interfaces are two-lane two-way streets that support Majorana current in both directions.  Increasingly complex surface termination patterns can support interface states of increasing complexity, but for such a state to support current flow, there must exist a closed interface-path that satisfies the above traffic rules.

\subsection{Majorana bound states}
\label{ssec:MBS}
\begin{figure}
\centerline{\resizebox{3.25in}{!}{\includegraphics{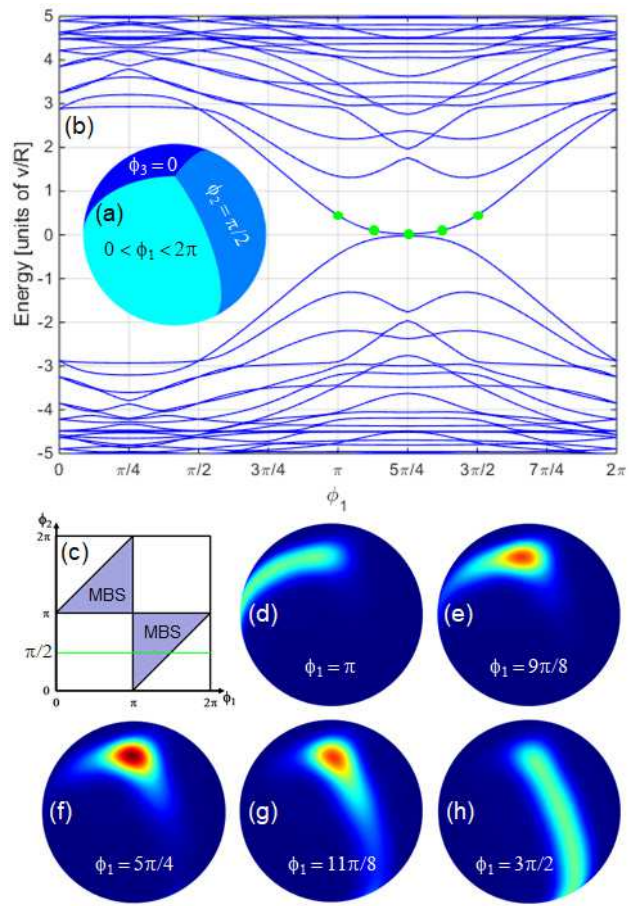}}}
\caption{Majorana bound states at poles.  (a) Spherical TI surface terminated by $s$-wave superconductors in three-region beach-ball pattern, with order parameters of the same magnitude, $|\Delta|$, but different phases: $0<\phi_1<2\pi$, $\phi_2 = \pi/2$, and $\phi_3 = 0$.  (b) Computed energy spectra as a function of phase $\phi_1$.  Note the development of near-zero-energy Majorana bound states (MBS) for $\pi < \phi_1 < 3\pi/2$, in agreement with the Fu-Kane \cite{fu2008superconducting} phase diagram shown in (c).  Quasiparticle density of the lowest positive energy state is plotted over the spherical surface in (d)-(h) for each of the five $\phi_1$ values marked in (b) by green dots.  These quasiparticle density snapshots illustrate the arrival and departure of the polar MBS across interfaces of phase difference $\pi$.  Results were computed via exact diagonalization up to $n_{\rm max} = 10$ (528 BdG states) for $|\Delta| = 4 v/R$ and $\mu = 0$.}
\label{fig:polarMBS}
\end{figure}

Fu and Kane \cite{fu2008superconducting} showed that superconductor trijunctions, TI surface terminations where three superconducting regions of different phase meet at a point, can support Majorana bound states (MBS) at the trijunction if the three phases are selected appropriately.  Such a configuration is easily realized within our spherical geometry.  The simple three-region beach-ball termination pattern depicted in Fig.~\ref{fig:polarMBS}(a) demonstrates this nicely.  Here we divide the spherical surface into three regions, separated by equally-spaced lines of longitude (meridians), where each region is terminated by $s$-wave superconductors with order parameters of the same magnitude but different phases, $\phi_1$, $\phi_2$, and $\phi_3$.  The interface lines meet at the north and south poles, resulting in two polar trijunctions.  We expect to find MBS at the poles for values of $\phi_1$ and $\phi_2$ within the shaded regions of the Fu-Kane \cite{fu2008superconducting} phase diagram, depicted in Fig.~\ref{fig:polarMBS}(c), where $\phi_3$ has been defined to be zero.

Our numerics provide a convenient way to explore the nature of this polar-trijunction system.  Consider a trace across the Fu-Kane phase diagram for $\phi_2 = \pi/2$, shown in green in the figure.  By varying $\phi_1$ from 0 to $2\pi$ in steps of $\pi/32$, we have calculated the energy spectra and eigenstates for every step along this trace.  The energy spectra are plotted versus $\phi_1$ in Fig.~\ref{fig:polarMBS}(b).  Note the two near-zero energy levels that develop as we enter the MBS region of the phase diagram ($\pi \leq \phi_1 \leq \frac{3\pi}{2}$).  These are the polar Majorana bound states.  Since our numerics compute not just energy levels but also eigenstates, it is straightforward and instructive to plot quasiparticle density corresponding to an MBS eigenstate as we move through the shaded MBS region of the phase diagram.  We do so in Fig.~\ref{fig:polarMBS}(d)-(h) for each of the points marked by green dots in Fig.~\ref{fig:polarMBS}(b): $\phi_1 = \pi, \frac{9\pi}{8}, \frac{5\pi}{4}, \frac{11\pi}{8}, \frac{3\pi}{2}$.  For $\phi_1 < \pi$, there are no MBS.  For $\phi_1 = \pi$, the interface between region 1 and region 3 becomes a $\pi$-junction since $\phi_1 - \phi_3 = \pi$.  As discussed in Sec.~\ref{ssec:propagating}, this allows Majorana current to flow along that interface, which it does as the MBS are created at the poles.  Fig.~\ref{fig:polarMBS}(d) shows the resulting quasiparticle density along the 1-3 interface meridian connecting the poles.  As $\phi_1$ increases toward $\frac{5\pi}{4}$, the center of the shaded MBS region, quasiparticle density becomes more concentrated at the poles, as seen in Fig.~\ref{fig:polarMBS}(d)-(f).  Then as $\phi_1$ increases further, beyond the center of the shaded MBS region, it spreads out once again, as seen in Fig.~\ref{fig:polarMBS}(f)-(h).  For $\phi_1 = \frac{3\pi}{2}$, the interface between region 1 and region 2 becomes a $\pi$-junction since $\phi_1 - \phi_2 = \pi$.  The resulting Majorana current along the 1-2 interface meridian allows the MBS to escape the poles and fuse/annihilate, as shown in Fig.~\ref{fig:polarMBS}(h).  For $\phi_1 > \frac{3\pi}{2}$, the MBS are gone once again.

\subsection{Mach-Zehnder interferometer}
\label{ssec:mach}
Now that we have seen how simple two-region and three-region termination patterns can be used to demonstrate the nature of propagating and bound Majorana states, let us consider a slightly more complex termination pattern that illustrates how propagating Majorana modes can be used to probe Majorana bound states.

\begin{figure}
\centerline{\resizebox{3.0in}{!}{\includegraphics{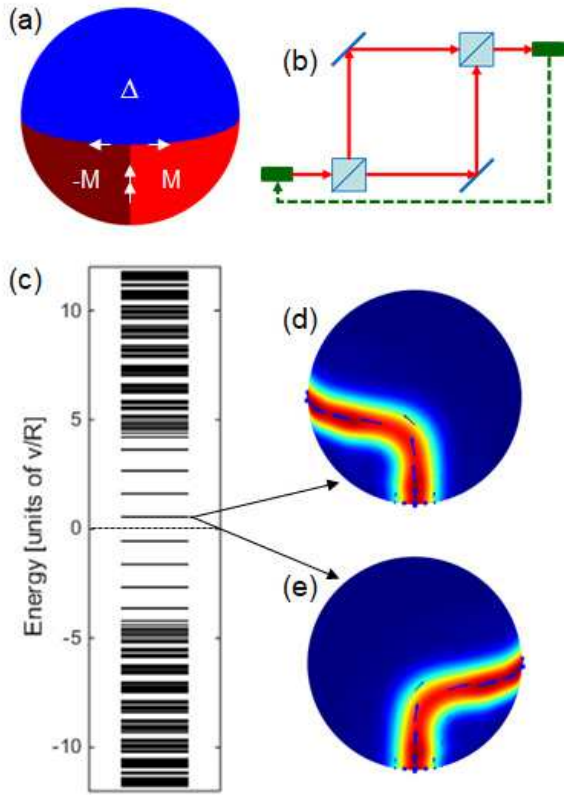}}}
\caption{Mach-Zehnder interferometer.  (a) Spherical TI surface terminated by $s$-wave superconductor $\Delta$ over the northern hemisphere, out-of-surface magnetization $M$ in the eastern half of the southern hemisphere, and into-surface magnetization $-M$ in the western half of the southern hemisphere.  Since two Majorana modes co-propagate northward from the south pole, split into two branches that propagate eastward and westward around the equator, and then merge again to co-propagate southward, this configuration is the Majorana analog of a Mach-Zehnder interferometer with its output fed back into its input, as depicted in (b). Since the northern hemisphere is uniform, this Mach-Zehnder is essentially empty.  Its energy spectrum (c) is therefore an evenly spaced ladder of doubly-degenerate levels, corresponding, respectively, to the modes that branch westward and eastward.  Plotted in (d) and (e) are the quasiparticle current density distributions for each of the lowest positive energy states.  Results were computed via exact diagonalization up to $n_{\rm max} = 10$ (528 BdG states) for $|\Delta| = |M| = 4 v/R$ and $\mu = 0$.}
\label{fig:machzehnder}
\end{figure}

We begin by constructing our probe, a spherical-TI-surface Majorana-current analog of a Mach-Zehnder interferometer \cite{fu2009probing,akhmerov2009electrically}.  Consider the three-region termination pattern depicted in Fig.~\ref{fig:machzehnder}(a).  The northern hemisphere is entirely $\Delta$-terminated, with uniform phase (for now).  The southern hemisphere is half terminated by $M$ and half terminated by $-M$, with the dividing interface running along a great circle segment from a point on the equator, through the south pole, back to the equator on the opposite side of the sphere.  As per the rules-of-thumb developed at the end of Sec.~\ref{ssec:propagating}, we expect current to flow in the directions indicated by the white arrows in the figure.  Two co-propagating modes are allowed along the $-M$/$M$ interface, and single propagating modes are allowed along the eastward-propagating $\Delta$/$M$ interface and the westward-propagating $\Delta$/$-M$ interface.  The resulting flow is that of two modes co-propagating northward from the south pole, splitting into two equatorial modes (one eastward, one westward), rejoining on the opposite side of the sphere, and co-propagating southward toward the south pole.  This situation is analogous to that of a Mach-Zehnder interferometer, with its output fed back into its input, as depicted in Fig.~\ref{fig:machzehnder}(b).  Since the two arms of this Mach-Zehnder wrap around the equator, it encloses the entire northern hemisphere and can be used as a probe thereof.  In the current configuration, the northern hemisphere is effectively empty, terminated by a single superconducting region of uniform phase.  Our numerics can be used to calculate the energy spectrum and eigenstates.  The resulting energy spectrum, plotted in Fig.~\ref{fig:machzehnder}(c), reveals interface states within the gap, corresponding to Majorana modes propagating through the Mach-Zehnder.  Each interface level is doubly degenerate, as there are two allowed modes.  The current patterns associated with the lowest positive energy states are plotted in Figs.~\ref{fig:machzehnder}(d) and (e).  The former depicts the mode that follows the westward branch while the latter depicts the mode that follows the eastward branch.  Since these two eigenstates are degenerate, linear combinations thereof, which follow both the westward and eastward branches, are also eigenstates, and could just as easily have been used to represent the propagating modes.

\begin{figure}
\centerline{\resizebox{3.25in}{!}{\includegraphics{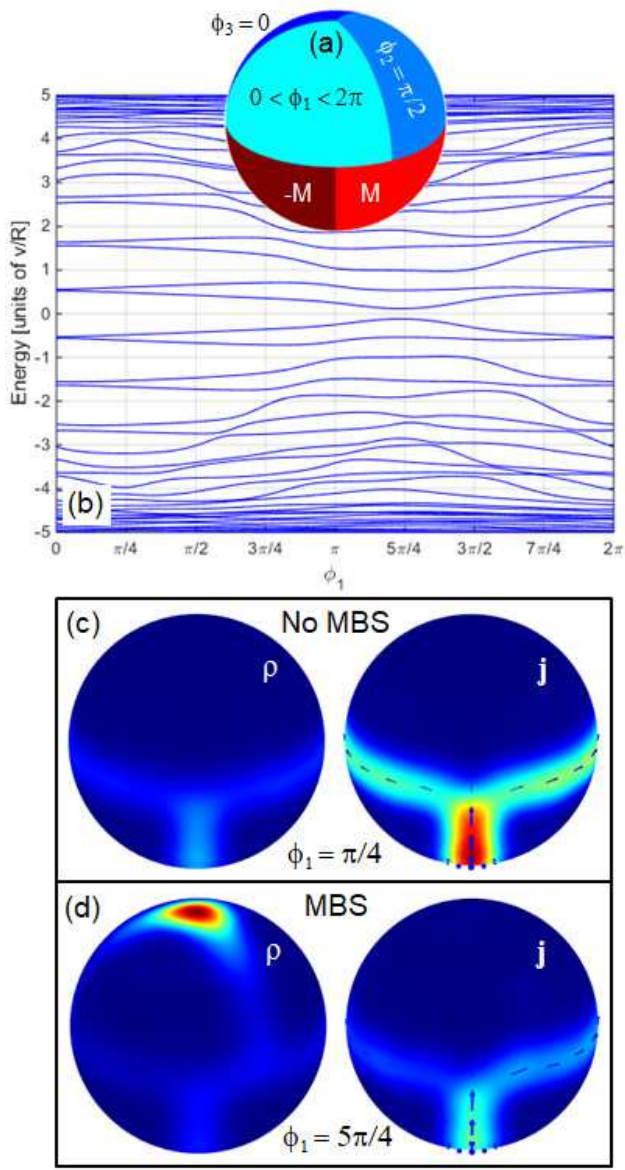}}}
\caption{Mach-Zehnder interferometer with trijunction at north pole.  (a) Five-region termination pattern consisting of Mach-Zehnder pattern from Fig.~\ref{fig:machzehnder} with northern-hemisphere superconductor replaced by three superconductor regions of different phase meeting at the north pole in a trijunction.  Energy spectra as a function of phase $\phi_1$ are plotted in (b).  Quasiparticle density $\rho$ and quasiparticle current density {\bf j} for the lowest positive energy state are plotted in (c) for $\phi_1=\pi/4$ and in (d) for $\phi_1=5\pi/4$.  In (c), the absence of an MBS at the north pole is accompanied by robust quasiparticle current in the Mach-Zehnder.  In (d), the presence of an MBS at the north pole is accompanied by a diminished quasiparticle current in the Mach-Zehnder.  Results were computed via exact diagonalization up to $n_{\rm max} = 10$ (528 BdG states) for $|\Delta| = |M| = 4 v/R$ and $\mu = 0$.}
\label{fig:machzehnderMBS}
\end{figure}

Now we can change the northern hemisphere a bit so our southern-hemisphere Mach-Zehnder has something interesting to probe.  Consider the five-region termination pattern depicted in Fig.~\ref{fig:machzehnderMBS}(a).  The $M$ and $-M$ regions in the southern hemisphere remain the same, but the single uniform-phase $\Delta$ region in the northern hemisphere has now been replaced by three equal-area $\Delta$ regions bounded by the equator and equally-spaced meridians that meet at the north pole.  The three order parameters have the same magnitude but different phases, $\phi_1$, $\phi_2$, and $\phi_3$.  Thus, while the equator continues to host the propagating modes that define the arms of the Mach-Zehnder, the north pole can support Majorana bound states if the phases are selected within the shaded region of the Fu-Kane phase diagram of Fig.~\ref{fig:polarMBS}(c).  Setting $\phi_3 = 0$ and $\phi_2 = \pi/2$, we can move along the green trace in the phase diagram by varying $\phi_1$ from 0 to $2\pi$ in steps of $\pi/32$.  Doing so, and calculating energy spectra at every step, yields the energy versus phase plots of Fig.~\ref{fig:machzehnderMBS}(b).  Note how these plots represent a hybridization of the phase-dependent spectra of the trijunction [Fig.~\ref{fig:polarMBS}(b)] and the phase-independent spectra of our Mach-Zehnder probe [Fig.~\ref{fig:machzehnder}(c)].  In addition to energy spectra, we compute, at every step, all 528 eigenstates, and can plot for each the associated quasiparticle density, $\rho$, and quasiparticle current density, ${\bf j}$, over the surface of the sphere.  For $\phi_1 = \pi/4$, the $\rho$ and ${\bf j}$ plots for the lowest-positive-energy eigenstate are shown in Fig.~\ref{fig:machzehnderMBS}(c).  Note the absence of quasiparticle density at the north pole.  This makes sense, since for $\phi_1 = \pi/4$, we are in the no-MBS (unshaded) part of the Fu-Kane phase diagram.  Note also that a robust Majorana current flows through the Mach-Zehnder at this value of $\phi_1$.  As we increase $\phi_1$ and enter the MBS (shaded) part of the Fu-Kane phase diagram, quasiparticle density builds up at the north pole.  This is evident from Fig.~\ref{fig:machzehnderMBS}(d), which contains the $\rho$ and ${\bf j}$ plots for $\phi_1 = 5\pi/4$.  Here, the quasiparticle density plot reveals a clear peak at the north pole.  The corresponding quasiparticle current density plot shows something interesting, a substantially diminished Majorana current throughout the Mach-Zehnder.

This diminished current is closely related to the change in energy level quantization that occurs in the presence of the MBS at the north pole, which acts as an effective $\pi$-flux between the arms of the interferometer. In the presence of this effective $\pi$-flux, our Mach-Zehnder setup is qualitatively similar (aside from the drain being fed back into the source) to the $\mathbb{Z}_2$ Majorana interferometer proposed by Fu and Kane~\cite{fu2009probing}. In Fig.~\ref{fig:machzehnderMBS}(b), which shows the energy spectrum as a function of the phase $\phi_1$ that controls the presence/absence of the north pole MBS, we see that for $\phi_1=0$, where the MBS is absent, the nearly doubly degenerate energy levels are approximately quantized as integer-plus-one-half multiples of the fundamental energy scale $v/R$. In the absence of the MBS, these two nearly degenerate modes correspond to equivalent linear combinations of eastward and westward propagating Majorana channels. As we increase $\phi_1$ and tune in an MBS at the north pole, these levels split such that, at $\phi_1 = 5\pi/4$, one of the energy levels is approximately quantized as integer multiples of $v/R$, while the other remains roughly unchanged. This implies that one of the modes is $\pi$ phase shifted with respect to the other, resulting in a net interference effect that is also reflected in the computed quasiparticle current.

\begin{figure}
\centerline{\resizebox{3.25in}{!}{\includegraphics{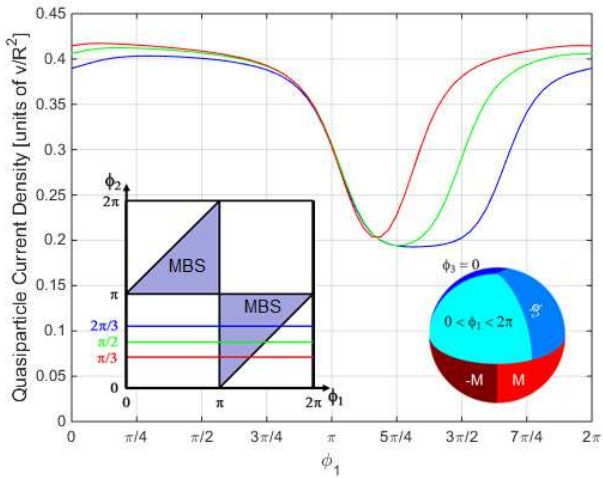}}}
\caption{Mach-Zehnder detection of a Majorana bound state.  For the lowest positive energy eigenstate associated with the depicted termination pattern (right inset), quasiparticle current density at the south pole is plotted as a function of phase $\phi_1$ for each of three traces across the Fu-Kane \cite{fu2008superconducting} phase diagram (left inset).  Note that south pole current is depressed by a factor of roughly $1/2$ in regions of phase space where there is a Majorana bound state at the north pole.  Hence, this configuration enables the detection of a north pole MBS via measurement of south pole current.  Results were computed via exact diagonalization up to $n_{\rm max} = 10$ (528 BdG states) for $|\Delta| = |M| = 4 v/R$ and $\mu = 0$.}
\label{fig:southpolecurrent}
\end{figure}

We can explore this effect further by computing quasiparticle current density right at the south pole as a function of phase $\phi_1$.  Results for the $\phi_2 = \pi/2$ case that we have been considering are plotted as the green curve in Fig.~\ref{fig:southpolecurrent}.  Note the dip in the south pole current by roughly one half, centered at $\phi_1 = 5\pi/4$, the center of the MBS region of the Fu-Kane phase diagram (see inset).  At half maximum, this dip extends from $\pi$ to $3\pi/2$, which is the parameter regime where the MBS is present at the north pole.  (The gradual nature of this dip is a consequence of the finite angular resolution of our numerics.)  The red and blue curves correspond to equivalent calculations of south pole current for the red trace ($\phi_2 = \pi/3$) and the blue trace ($\phi_2 = 2\pi/3$) in the inset phase diagram.  Once again, there is a dip in the south pole current by roughly one half for $\phi_1$ within the range where the north pole hosts an MBS.  This range is narrower for the red curve because the red trace crosses a narrower slice of the MBS region of the phase diagram, and is wider for the blue curve because the blue trace crosses a wider slice of the MBS region.  In the presence of a north pole MBS, there is a phase difference between the eastward and westward propagating arms of the Mach-Zehnder.  The resulting interference leads to a diminished south pole current whenever there is an MBS at the north pole.  Hence, our southern hemisphere Mach-Zehnder functions as a detector of Majorana bound states at the north pole.

\section{Conclusions}
\label{sec:conclusions}
In this work, we numerically solved the Bogoliubov-de Gennes equation for a spherical topological insulator surface characterized by a piecewise-uniform proximity-induced superconducting order parameter and out-of-surface magnetization.  The interface of two independent gapping terms forms a domain wall that carries Majorana excitations.  By using different patterning of these gapping terms, one can implement both bound-state and propagating Majorana modes.  We developed a computational framework that allows us to prototype a wide variety of Majorana circuits simply by specifying the surface termination pattern in an input file.  Using exact diagonalization, we computed the energy spectra and eigenstates of Majorana excitations at domain wall interfaces and trijunctions.  For each computed eigenstate, our numerics generates plots of quasiparticle density and quasiparticle current density over the surface of the sphere, some of which were presented in this paper.  We used this utility to investigate the non-Abelian nature of Majorana excitations by probing their braiding properties.  To this end, we implemented a Mach-Zehnder interferometer circuit with equatorial interferometer arms surrounding a polar trijunction.  By manipulating the phases of the superconducting order parameters, we were able to tune in a Majorana bound state that acts as a $\pi$-flux within the area of the interferometer.  We computed the interference signatures in the quasiparticle current density and found results consistent with the Fu-Kane trijunction phase diagram~\cite{fu2008superconducting}.

The cases considered in this paper are just a sample of the many surface termination patterns that can be probed using the computational utility developed herein.  Possibilities for studying increasingly complex Majorana circuits are limited only by computation time and imagination.  Since the large radius limit of the spherical geometry reproduces flat manifold physics~\cite{imura2012spherical}, our results can be extended to realistic experimental setups.

The main strength of our work lies in the numerical flexibility of our computational framework, which will make it possible, in the future, to add realistic effects of disorder and interactions, and study how they may alter transport signatures in experiments.  The role of disorder in (unterminated) TI surface states has recently been studied in Ref.~\onlinecite{durst2016disorder}.  Extending the framework developed in this paper to include disorder will allow us to study its effects on the braiding and transport signatures of Majorana circuits, effects which we expect to play a key role in understanding experimental probes of the non-Abelian nature of Majorana excitations.  Our formalism is also well-suited to include interaction effects~\cite{neupert2015interacting} on these braiding and transport signatures, providing an additional path forward for future work.  Another interesting direction will be to consider the effect of the proximity of unconventional and topological superconductors on the surface states, and to study the interface excitations arising therefrom.

\begin{acknowledgments}
A.C.D. is grateful to B. Burrington and G. C. Levine for helpful discussions.  He was supported by funds provided by Hofstra University, including a Faculty Research and Development Grant (FRDG), a Presidential Research Award Program (PRAP) grant, and faculty startup funding.  His work was performed in part as a KITP Scholar at the Kavli Institute for Theoretical Physics, supported by the National Science Foundation under Grant No.\ NSF PHY-1748958.  S.G. was supported by startup funds provided by a 21st Century Foundation grant at CCNY-CUNY.  His work was performed in part at the Aspen Center for Physics, which is supported by the National Science Foundation under Grant No.\ NSF PHY-1607611.
\end{acknowledgments}

\end{document}